\documentclass[graybox]{svmult}

\usepackage{hyperref}
\usepackage{graphicx}
\usepackage{epstopdf}
\usepackage{mathptmx}       % selects Times Roman as basic font
\usepackage{helvet}         % selects Helvetica as sans-serif font
\usepackage{courier}        % selects Courier as typewriter font
\usepackage{type1cm}        % activate if the above 3 fonts are
\usepackage{makeidx}         % allows index generation
\usepackage{multicol}        % used for the two-column index
\usepackage[bottom]{footmisc}% places footnotes at page bottom

\makeindex             % used for the subject index
\begin{document}

\title*{Dilute Fermi and Bose Gases}
\author{Subir Sachdev}
\institute{Subir Sachdev \at Department of Physics, Harvard University, Cambridge MA 02138, USA\\ \email{sachdev@physics.harvard.edu}}
\maketitle

\abstract*{We give a unified perspective on the properties of a variety of quantum liquids using the theory
of quantum phase transitions. A central role is played by a zero density quantum critical point which is argued to control
the properties of the dilute gas. An exact renormalization group analysis of such quantum critical points
leads to a computation of the universal properties of the dilute Bose gas and the spinful Fermi gas near a Feshbach resonance.
}

\abstract{I give a unified perspective on the properties of a variety of quantum liquids using the theory
of quantum phase transitions. A central role is played by a zero density quantum critical point which is argued to control
the properties of the dilute gas. An exact renormalization group analysis of such quantum critical points
leads to a computation of the universal properties of the dilute Bose gas and the spinful Fermi gas near a Feshbach resonance.
}

\section{Introduction}
\label{sec:intro}

This article is adapted from Chapter 16 of \href{http://sachdev.physics.harvard.edu/qptweb/toc.html}{{\em Quantum Phase Transitions}, 2nd edition},
Cambridge University Press.

It is not conventional to think of dilute quantum liquids as being in the vicinity
of a quantum phase transition. However, there is a simple sense in which they
are, although there is often no broken symmetry or order parameter associated 
with this quantum phase transition. 
We shall show below that the perspective of such a quantum phase transition
allows a unified and efficient description of the universal properties of quantum liquids.

Stated most generally, consider a quantum liquid with a 
global ${\rm U}(1)$ symmetry.
We shall be particularly
interested in the behavior of the conserved density, generically
denoted by $Q$ (usually the particle number), associated with this symmetry.  
The quantum phase transition is between two phases with
a specific $T=0$ behavior in the expectation value of $Q$.
In one of the phases, $\langle Q \rangle$ is pinned precisely at
a quantized value (often zero) and does not vary as microscopic parameters
are varied. This quantization ends at
the quantum critical point with a discontinuity in the
derivative of $\langle Q \rangle$ with respect to the tuning
parameter (usually the chemical potential), 
and $\langle Q \rangle$ varies smoothly in the other
phase; there is no discontinuity in the value of $\langle Q
\rangle$, however.

The most familiar model exhibiting such a quantum phase transition
is the dilute Bose gas. We express its coherent state partition function, $Z_B$,
in terms of complex field $\Psi_B (x, \tau)$, where $x$ is a $d$-dimensional
spatial co-ordinate and $\tau$ is imaginary time:
\begin{eqnarray}
 Z_B &=& ~\int {\cal D} \Psi_B (x,\tau) 
\exp\left( - ~\int_0^{1/T} d \tau ~\int d^d x {\cal L}_{B} \right), \nonumber \\
{\cal L}_{B} &=&  \Psi_B^{\ast} ~\frac{\partial \Psi_B}{\partial \tau}
 + ~\frac{1}{2m} \left| \nabla \Psi_B \right|^2 -\mu |\Psi_B|^2 + \frac{u_0}{2} |\Psi_B|^4.
\label{xx0}
\end{eqnarray}
We can identify the charge $Q$ with the boson density $\Psi_B^{\ast} \Psi_B$ 
\begin{equation}
\langle Q \rangle
 = - \frac{\partial {\cal F}_B}{\partial \mu} = \langle |\Psi_B|^2 \rangle,
\label{xx0a}
\end{equation}
with ${\cal F}_B = - (T/V) \ln Z_B$.
The quantum critical point is precisely at $\mu = 0$ and $T=0$, and 
there are {\em no} fluctuation corrections to this
location from the terms in ${\cal L}_B$.
So at $T=0$, $ \langle Q \rangle$ takes the quantized value
$\langle Q \rangle = 0$ for $\mu < 0$, and $\langle Q \rangle  > 0$
for $\mu > 0$; we will describe the nature of the onset
at $\mu=0$ and finite-$T$ crossovers in its vicinity. 

Actually, we will begin our analysis in Section~\ref{sec:fermigas}
by a model simpler than $Z_B$, which
displays a quantum phase transition with the same behavior in a conserved ${\rm U}(1)$
density $\langle Q \rangle$ and has many
similarities in its physical properties. The model is exactly solvable and
is expressed in terms
of a continuum canonical spinless fermion field $\Psi_F$; its partition function
is
\begin{eqnarray}
Z_F &=& ~\int {\cal D} \Psi_F (x,\tau) 
\exp\left( - ~\int_0^{1/T} d \tau ~\int d^d x {\cal L}_{F} \right),
 \nonumber \\
{\cal L}_{F} &=&  \Psi_F^{\ast} ~\frac{\partial \Psi_F}{\partial \tau}
 + ~\frac{1}{2m} \left| \nabla \Psi_F \right|^2 -\mu |\Psi_F|^2 .
\label{xx0b}
\end{eqnarray}
${\cal L}_F$ is just a free field theory. Like $Z_B$,
$Z_F$ has a quantum critical point at $\mu=0$, $T=0$ and we will discuss
its properties; in particular, we will show that all
possible fermionic nonlinearities are irrelevant near it.
The reader should not be misled by the apparently trivial
nature of the model in (\ref{xx0b}); using the theory of quantum phase transitions
to understand free fermions might seem like technological overkill. We will see
that $Z_F$ exhibits crossovers that are quite similar to those near far more
complicated quantum critical points, and observing them in this simple
context leads to considerable insight.

In general spatial dimension, $d$, the continuum theories $Z_B$ and $Z_F$
have different, though closely related, universal properties. However,
we will argue that the quantum critical points of
these theories are {\em exactly} equivalent in $d=1$. We will see that
the bosonic theory $Z_B$ is strongly coupled in $d=1$, and
will note
compelling evidence that the solvable fermionic theory $Z_F$ is its exactly universal
solution in the vicinity of the $\mu=0$, $T=0$ quantum critical point. This equivalence extends
to observable operators in both theories, and allows exact computation of a number of universal
properties of $Z_B$ in $d=1$.

Our last main topic will be a discussion of the
dilute spinful Fermi gas in Section~\ref{sec:feshbach}. This generalizes $Z_F$ to a 
spin $S=1/2$ fermion $\Psi_{F\sigma}$, with $\sigma = \uparrow, \downarrow$. 
Now Fermi statistics do allow a contact quartic interaction, and so we have
\begin{eqnarray}
Z_{Fs} &=& ~\int {\cal D} \Psi_{F\uparrow} (x,\tau) {\cal D} \Psi_{F\downarrow} (x,\tau)
\exp\left( - ~\int_0^{1/T} d \tau ~\int d^d x \, {\cal L}_{Fs} \right),
 \nonumber \\
{\cal L}_{Fs} &=&  \Psi_{F\sigma}^{\ast} ~\frac{\partial \Psi_{F\sigma}}{\partial \tau}
 + ~\frac{1}{2m} \left| \nabla \Psi_{F\sigma} \right|^2 -\mu |\Psi_{F\sigma}|^2 + u_0 \Psi_{F\uparrow}^\ast
 \Psi_{F \downarrow}^\ast \Psi_{F\downarrow} \Psi_{F \uparrow}.
\label{fesh1}
\end{eqnarray}
This theory conserves fermion number, and has a phase transition as a function of 
increasing $\mu$ from a state with fermion number 0 to a state with non-zero fermion density.
However, unlike the above two cases of $Z_B$ and $Z_F$, the transition is not
always at $\mu=0$. The problem defined in (\ref{fesh1}) has recently found remarkable experimental
applications in the study of ultracold gases of fermionic atoms. These experiments are aslo
able to tune the value of the interaction $u_0$ over a wide range of values, extended from
repulsive to attractive. For the attractive case, the two-particle scattering amplitude has a 
Feshbach resonance where the scattering length diverges, and we obtain the unitarity limit.
We will see that this
Feshbach resonance plays a crucial role in the phase transition obtained by changing $\mu$,
and leads to a rich phase diagram of the so-called ``unitary Fermi gas''.

Our treatment of $Z_{Fs}$ in the experimental important case of $d=3$ will show that
it defines a strongly coupled field theory in the vicinity of the Feshbach resonance for attractive
interactions. It therefore pays to find alternative formulations of this regime of the unitary Fermi gas. 
One powerful
approach is to promote the two fermion bound state to a separate canonical Bose field.
This yields a model, $Z_{FB}$ with both elementary fermions and bosons ; {\em i.e.\/} it is a combination
of $Z_B$ and $Z_{Fs}$ with interactions between the fermions and bosons. We will define
$Z_{FB}$ in Section~\ref{sec:feshbach}, and use it to obtain a number of experimentally relevant
results for the unitary Fermi gas.

Section~\ref{sec:fermigas} will present a thorough discussion of the
universal properties
of $Z_F$. This will be followed by an analysis of $Z_B$ in Section~\ref{sec:xx3}, 
where we will use renormalization group methods to obtain perturbative predictions
for universal properties. 
The spinful Fermi gas will be discussed in Section~\ref{sec:feshbach}.

\section{The Dilute Spinless Fermi Gas}
\label{sec:fermigas}

This section will study the properties of $Z_F$ in the vicinity of its
$\mu=0$, $T=0$ quantum critical point. As $Z_F$ is a simple free field theory,
all results can be obtained exactly and are not particularly profound in
themselves. Our main purpose is to show how the results are interpreted in a
scaling perspective and to obtain general lessons on the nature of crossovers
at $T>0$. 

First, let us review the basic nature of the quantum critical point at $T=0$.
A useful diagnostic for this is the conserved density $Q$, which in the present
model we identify as $\Psi_F^{\dagger} \Psi_F$. As a function of the
tuning parameter $\mu$, this quantity has a critical singularity at
$\mu=0$:
\begin{equation}
\big\langle \Psi^{\dagger}_F \Psi_F \big\rangle =  \left\{
\begin{array}{c@{\quad}c}
(S_d /d) (2 m \mu)^{d/2}, & \mu > 0, \\
0, & \mu < 0,
\end{array}
\right.
\label{xx5}
\end{equation}
where the phase space factor $S_d = 2 /[\Gamma(d/2) (4 \pi)^{d/2}]$.

We now proceed to a scaling analysis. Notice that at the quantum critical
point $\mu = 0$, $T=0$, the theory ${\cal L}_F$ is invariant under
the scaling transformations:
\begin{eqnarray}
x' &=& x e^{-\ell}, \nonumber \\
\tau' &=& \tau e^{-z\ell}, \label{xx7}\\
\Psi'_F &=& \Psi_F e^{d \ell/2}, \nonumber 
\end{eqnarray}
provided we make the choice of the dynamic
exponent
\begin{equation}
z = 2.
\label{xx7a}
\end{equation}
The parameter $m$ is assumed to remain invariant under the rescaling,
and its role is simply to ensure that the relative physical dimensions
of space and time are compatible.
The transformation (\ref{xx7}) also identifies the scaling
dimension
\begin{equation}
\mbox{dim} [ \Psi_F ] = d/2.
\label{xx7b}
\end{equation}

Now turning on a nonzero $\mu$, it is easy to see that $\mu$ is a relevant perturbation
with
\begin{equation}
\mbox{dim} [\mu ] = 2.
\label{xx7c}
\end{equation}
There will be no other relevant perturbations at this quantum critical point,
and so we have for the correlation length exponent
\begin{equation}
\nu = 1/2.
\label{xx7d}
\end{equation}

We can now examine the consequences of adding interactions to ${\cal L}_F$.
A contact interaction such as $\int d x (\Psi^{\dagger}_F (x) \Psi_F(x))^2 $ vanishes because
of the fermion anticommutation relation. (A contact interaction is however
permitted for a spin-1/2 Fermi gas and will be discussed in Section~\ref{sec:feshbach}) The
 simplest allowed term for the spinless Fermi gas
 is
\begin{equation}
{\cal L}_1 = \lambda \big( \Psi_F^{\dagger}(x,\tau) \nabla \Psi_F^{\dagger} (x, \tau)
\Psi_F (x, \tau) \nabla \Psi_F (x, \tau) \big),
\label{xx7e}
\end{equation}
where $\lambda$ is a coupling constant measuring the strength of the interaction.
However, a simple analysis  shows that
\begin{equation}
\mbox{dim}[\lambda] = -d.
\label{xx7f}
\end{equation}
This is negative and so $\lambda$ is irrelevant and can be neglected in the computation
of universal crossovers near the point $\mu=T=0$. In particular, it will modify
the result (\ref{xx5}) only by contributions that are higher order in $\mu$.

\begin{figure}[t]
%\epsfscale330
%\epsfxsize=140pt
\centerline{\includegraphics[width=3.5in]{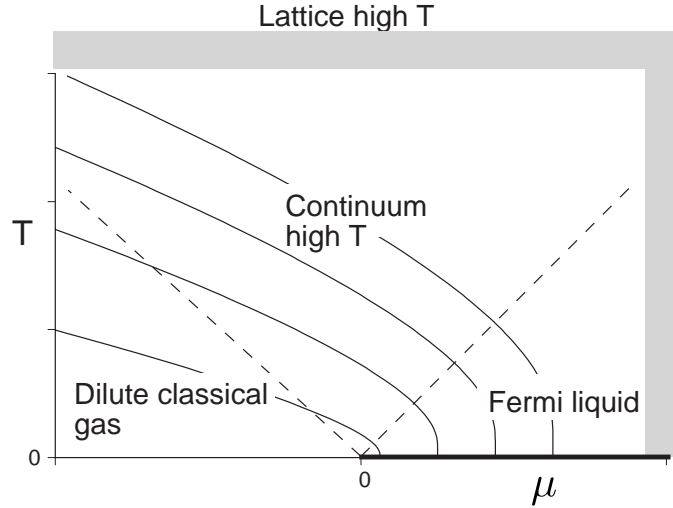}}
\caption{Phase diagram of the dilute Fermi gas $Z_F$ (Eqn. (\protect\ref{xx0b}))
as a function of the chemical potential $\mu$ and the temperature $T$.
The regions are separated by crossovers denoted by dashed lines, and their physical properties %
are discussed in the text.
The full lines are contours of equal density, with higher densities above lower densities;
the zero density line is $\mu <0$, $T=0$.
The line $\mu > 0$, $T=0$ is a line of $z=1$ critical points
that controls the longest scale properties of the low-$T$ Fermi liquid region.
The critical end point $\mu=0$, $T=0$ has $z=2$ and controls global structure of
the phase diagram. In $d=1$, the Fermi liquid is more appropriately labeled
a Tomonaga--Luttinger liquid.
The shaded region marks the boundary of applicability
of the continuum theory and occurs at $\mu, T \sim w$.}
\label{xxf2}
\end{figure}

Turning to nonzero temperatures, we can  write down scaling forms.
Let us define the fermion Green's function
\begin{equation}
G_F (x, t) = \big\langle \Psi_F (x,t) \Psi_F^{\dagger} (0,0) \big\rangle;
\label{xx7g}
\end{equation}
then the scaling dimensions above imply that it
satisfies
\begin{equation}
G_F (x, t)
= \left(2 m T \right)^{d/2} \Phi_{G_F} \left(
(2 m  T)^{1/2} x , Tt , \frac{\mu}{T} \right),
\label{xx10}
\end{equation}
where
$\Phi_{G_F}$ is a fully universal scaling function.
For this particularly simple theory ${\cal L}_F$ we can of course
obtain the result for $G_F$ in closed form:
\begin{equation}
G_F ( x, t) = \int \frac{d^d k}{(2 \pi)^d}
\frac{e^{i k x -i(k^2/(2m) - \mu) t} }{1 + e^{-(k^2 / (2m) -\mu)/T}},
\label{xx11}
\end{equation}
and it is easy to verify that this obeys the scaling form (\ref{xx10}).
Similarly the free energy ${\cal F}_F$ has scaling dimension $d+z$, and we have
\begin{equation}
{\cal F}_F = T^{d/2+1} \Phi_{{\cal F}_F} \left( \frac{\mu}{T} \right)
\label{xx10a}
\end{equation}
with
$\Phi_{{\cal F}_F}$ a universal scaling function; the explicit result is,
of course,
\begin{equation}
{\cal F}_F = - \int \frac{d^d k}{(2 \pi)^d} \ln \big(
1 + e^{(\mu - k^2 / (2 m))/T}\big),
\label{xx12}
\end{equation}
which clearly obeys (\ref{xx10a}).
The crossover behavior of the fermion density
\begin{equation}
\langle Q \rangle
= \big\langle \Psi_F^{\dagger} \Psi_F\big\rangle = - \frac{\partial {\cal F}_F}{\partial \mu}
\label{xx12z}
\end{equation}
 follows by taking the appropriate derivative of
the free energy.
Examination of these results leads to the 
crossover phase diagram of Fig.~\ref{xxf2}.  
We will examine each of the regions of the phase diagram in turn,
beginning with the two low-temperature regions.

\subsection{Dilute Classical Gas, $k_B T \ll |\mu|$, $\mu < 0$}
\label{sec:xxcg}

The ground state for $\mu < 0$ is the vacuum with no particles. Turning on a
nonzero temperature produces particles with a small nonzero density $\sim\!\!e^{-|\mu|/T}$.
The de Broglie wavelength of the particles is of order $T^{-1/2}$, which is significantly
smaller than the mean spacing between the particles, which diverges as $e^{|\mu|/dT}$ as $
T \rightarrow 0$. This implies that the particles behave semiclassically.
To leading order from  (\ref{xx11}),
the fermion Green's function is simply the Feynman propagator of a single particle
\begin{equation}
G_F (x,t) = \left( \frac{m}{2 \pi i t} \right)^{d/2}
\exp \left( - \frac{i m x^2}{2 t} \right),
\label{xx8}
\end{equation}
and the exclusion of states from the
other particles has only an exponentially small effect.
Notice that $G_F$ is independent of $\mu$ and $T$ and (\ref{xx8}) is the exact result
for $\mu=T=0$.
The free energy, from (\ref{xx10a}) and (\ref{xx12}), is that of a classical
Boltzmann gas
\begin{equation}
{\cal F}_F = - \left( \frac{m T}{2 \pi} \right)^{d/2} e^{- |\mu|/T}.
\label{xx11a}
\end{equation}

\subsection{Fermi Liquid, $k_B T \ll \mu$, $\mu > 0$}
\label{sec:xxflt}

The behavior in this regime is quite complex and rich. As we will see, and as
noted in Fig.~\ref{xxf2}, the line $\mu > 0$, $T=0$ is itself a line
of quantum critical points.
The interplay between these critical points
and those of the $\mu=0$, $T=0$ critical end point is displayed quite
instructively in the exact results for $G_F$ and is worth examining in detail.
It must be noted that the scaling dimensions and critical exponents
of these two sets of critical points need not (and indeed will not) be the same.
The 
behavior of the $\mu > 0$, $T=0$ critical line emerges as a 
particular scaling limit of the global scaling
functions of the $\mu =0$, $T=0$ critical end point. Thus the latter scaling functions are 
globally valid everywhere in Fig.~\ref{xxf2}, and describe the physics of all its regimes.

First it can be argued, for example, by studying asymptotics of the integral in
(\ref{xx11}), that for very short times or distances, the correlators do not notice the
consequences of other particles present because of a
nonzero $T$ or $\mu$ and are therefore given by the single-particle propagator,
which is the
$T=\mu=0$
result in (\ref{xx8}). More precisely we have
\begin{equation}
\mbox{$G(x,t)$ is given by (\ref{xx8}) for}~|x| \ll \left( 2 m \mu
\right)^{-1/2},\quad|t| \ll \frac{1}{\mu}.
\label{xx12a}
\end{equation}
With increasing $x$ or $t$, the restrictions in (\ref{xx12a}) are eventually violated
and the consequences of the presence of other particles,
resulting from
a nonzero $\mu$, become apparent. Notice that because $\mu$ is
much larger than $T$, it is the first energy scale to be noticed, and as a first
approximation to understand the behavior at larger $x$ we may ignore the effects of $T$.

Let us therefore discuss the ground state for $\mu > 0$. It consists of a filled Fermi
sea of particles (a Fermi liquid) with momenta $k < k_F = (2 m \mu)^{1/2}$. An
important  property of the this state is that it permits excitations at arbitrarily low
energies
(i.e., it is {\em gapless}). These low energy excitations correspond to changes in
occupation number of fermions arbitrarily close to $k_F$. As a consequence of these gapless
excitations,  the points
$\mu > 0$ ($T=0$) form a line of quantum critical points, as claimed earlier.
We will now derive the continuum field theory associated with this line of
critical points. We are interested here only in $x$
and $t$ values that violate the constraints in (\ref{xx12a}), and so in
occupation of states with momenta near $\pm k_F$. So let us parameterize, in $d=1$,
\begin{equation}
\Psi (x, \tau) = e^{ik_F x} \Psi_R (x, \tau) +
e^{-ik_F x} \Psi_L (x, \tau),
\label{xx12b}
\end{equation}
where $\Psi_{R,L}$ describe right- and left-moving fermions and are fields that vary
slowly on spatial scales $\sim\!\!1/k_F = (1/{ 2 m \mu})^{1/2}$ and temporal scales
$\sim\!\!1 /\mu$; most of the results discussed below hold, with small
modifications, in all $d$. Inserting the above parameterization in
${\cal L}_F$, and keeping only terms lowest order in spatial gradients, we obtain the
``effective'' Lagrangean for the Fermi liquid region, ${\cal L}_{FL}$
in $d=1$:
\begin{equation}
{\cal L}_{FL} =
\Psi_R^{\dagger} \left( \frac{\partial}{\partial \tau}
- i v_F \frac{\partial}{\partial x} \right) \Psi_R
+  \Psi_L^{\dagger} \left( \frac{\partial}{\partial \tau}
+ i v_F \frac{\partial}{\partial x} \right) \Psi_L,
\label{xx13}
\end{equation}
where $v_F =  k_F / m = (2 \mu/m)^{1/2} $ is the Fermi velocity.
Now notice that ${\cal L}_{FL}$ is
invariant under a  scaling transformation, which is rather different from (\ref{xx7}) for
the $\mu=0$, $T=0$ quantum critical point:
\begin{equation}
\begin{array}{rcl}
x' &=& x e^{-\ell},\\[4pt]
\tau' &=& \tau e^{-\ell}, \\[4pt]
\Psi_{R,L}' (x', \tau') &=& \Psi_{R,L} (x, \tau) e^{\ell /2}, \\[4pt]
v_F' &=& v_F.
\label{xx14}
\end{array}
\end{equation}
The above results imply
\begin{equation}
z=1,
\label{xx14a}
\end{equation}
unlike $z=2$ (Eqn. (\ref{xx7a})) at the $\mu=0$ critical point, and
\begin{equation}
\mbox{dim}[\Psi_{R,L}] = 1/2,
\label{xx14b}
\end{equation}
which actually holds for all $d$ and therefore differs from
(\ref{xx7b}). Further notice that $v_F$, and therefore $\mu$, are  {\em invariant} under
rescaling, unlike (\ref{xx7c}) at the $\mu=0$ critical point. Thus
$v_F$  plays a role rather analogous to that of
$m$ at the
$\mu=0$ critical point: It is simply the physical units of spatial and length scales.
The transformations (\ref{xx14}) show that ${\cal L}_{LF}$ is scale invariant for each
value of $\mu$, and we therefore have a line of quantum critical points as claimed earlier.
It should also be emphasized that the scaling dimension of
interactions such as $\lambda$ will
 also change; in particular not all interactions are irrelevant about the $\mu\neq 0$
critical points. These new interactions are, however, small in magnitude provided
$\mu$ is small (i.e., provided we are within the domain of validity
of the global scaling forms (\ref{xx10}) and (\ref{xx10a}), and so we will neglect them here.
Their main consequence is to change the scaling dimension of certain operators,
but they preserve the relativistic and conformal invariance of ${\cal L}_{FL}$.
This more general theory of $d=1$ fermions is the Tomonaga--Luttinger liquid.

\subsection{High-${T}$ Limit, $k_B T \gg |\mu|$}
\label{sec:xxhight}

This is the last, and in many ways the most interesting, region of Fig.~\ref{xxf2}.  
Now $ T $ is the most important energy scale controlling the deviation from the
$\mu=0$, $T=0$ quantum critical point, and the properties will therefore
have some similarities to the ``quantum critical region'' of other strongly interacting models
\cite{book}.
It should be emphasized that while the value of
$ T$ is significantly larger than $|\mu|$, it cannot be so large that it exceeds
the limits of
applicability for the continuum action ${\cal L}_F$.
If we imagine that ${\cal L}_F$ was obtained from a model of lattice
fermions with bandwidth $w$, then we must have $T \ll w$.

We discuss first the behavior of the
fermion density. In the high-$T$ limit of the continuum theory ${\cal L}_F$,
$|\mu| \ll  T \ll w$, we have from (\ref{xx12}) and (\ref{xx12z}) the universal result
\begin{eqnarray}
\big\langle \Psi^{\dagger}_F \Psi_F \big\rangle &=&  \left({2 m  T} \right)^{d/2}
\int \frac{d^d y}{(2 \pi)^d} \frac{1}{e^{y^2} + 1} \nonumber \\[5pt]
&=& \left( {2 m  T} \right)^{d/2} \zeta (d/2) \frac{ ( 1 - 2^{d/2}
)}{(4 \pi)^{d/2}}.
\label{xx16}
\end{eqnarray}
This density implies an interparticle spacing that is of order the
de Broglie\index{de Broglie wavelength}
wavelength $= (1/2 m  T)^{1/2}$. Hence thermal and quantum effects are
to be equally important, and neither dominate.

For completeness,
let us also consider the fermion density for $T \gg w$ (the region above the shaded region
in Fig.~\ref{xxf2}), to illustrate the limitations on the continuum description discussed above. %
Now
the result depends upon the details of the nonuniversal fermion dispersion; on a hypercubic
lattice with dispersion $\epsilon_k-\mu$, we obtain
\begin{eqnarray}
\big\langle \Psi^{\dagger}_F \Psi_F \big\rangle
&=& \int_{-\pi/a}^{\pi/a}\frac{d^d k}{(2 \pi)^d} \frac{1}{e^{(\varepsilon_k - \mu)/T} + 1}
\nonumber\\[5pt] 
&=& \frac{1}{2 a^d} -\frac{1}{4T}  \int_{-\pi/a}^{\pi/a}\frac{d^d k}{(2 \pi)^d}
(\varepsilon_k -
\mu)  + {\cal O}(1/T^{2} ) .
\end{eqnarray}
The limits on the integration, which extend from $-\pi/a$ to $\pi/a$ for each momentum
component, had previously been sent to infinity in the continuum limit $a \rightarrow 0$.
In the presence of lattice cutoff, we are able to make a naive expansion of the integrand in
powers of $1/T$, and the result therefore only contains negative integer powers of $T$.
Contrast this with the universal continuum result (\ref{xx16}) where we had noninteger
powers of $T$ dependent upon the scaling dimension of
$\Psi$.

We  return to the universal high-$T$ region, $|\mu| \ll  T \ll w$,
and describe the behavior of the fermionic Green's function $G_F$,
given in (\ref{xx11}).
At the shortest scales we again have the free quantum particle behavior of the
$\mu = 0$, $T =0$ critical point:
\begin{eqnarray}
&\mbox{$G_F (x,t)$ is given by (\ref{xx8}) for}~|x| \ll \left( { 2 m  T}
\right)^{-1/2}, |t| \ll \,\frac{1}{ T}.
\label{xx17}
\end{eqnarray}
Notice that the limits on $x$ and $t$ in (\ref{xx17}) are different from those in
(\ref{xx12a}), in that they are determined by $ T$ and not $\mu$.
At larger $|x|$ or $t$ the presence of the other thermally excited particles
becomes apparent, and $G_F$ crosses over to a novel behavior
characteristic of the high-$T$ region.
We illustrate this by looking at the large-$x$ asymptotics of the equal-time $G$
in $d=1$ (other $d$ are quite similar):
\begin{equation}
G_F (x, 0) = \int \frac{dk}{2 \pi} \frac{e^{ikx}}{
1 + e^{- k^2 /2 m  T}}.
\end{equation}
For large $x$ this can be evaluated by a contour integration, which picks up contributions
from the poles at which the denominator vanishes in the complex $k$ plane. The dominant
contributions come from the poles closest to the real axis, and give the leading result
\begin{eqnarray}
&& G_F (|x| \rightarrow \infty, 0) =
- \left( \,\frac{\pi ^2}{2 m  T} \right)^{1/2}
\exp\left( - ( 1- i) \left( {m \pi  T}\right)^{1/2} x\right).
\label{xx18}
\end{eqnarray}
Thermal effects therefore lead to an exponential decay of equal-time correlations,
with a correlation length $\xi = \left({m \pi  T} \right)^{-1/2}$.
Notice that the $T$ dependence is precisely that expected from the exponent
$z=2$ associated with the $\mu=0$ quantum critical point and the general scaling relation
$\xi \sim T^{-1/z}$. The additional oscillatory term in (\ref{xx18}) is a reminder that
quantum effects are still present at the scale $\xi$, which is clearly of order the de
Broglie wavelength of the particles.

\section{The Dilute Bose Gas}
\label{sec:xx3}

This section will study the universal properties quantum phase transition
of the dilute Bose gas
model $Z_B$ in (\ref{xx0}) in general dimensions. We will begin with a simple
scaling analysis that will show that $d=2$ is the upper-critical dimension.
The first subsection will analyze the case $d<2$ in some more detail,
while the next subsection will consider the somewhat different properties in $d=3$. Some of the results of this section were also obtained by Kolomeisky and Straley 
\cite{KS1,KS2}.

We begin with the analog of the simple scaling considerations presented at the
beginning of Section~\ref{sec:fermigas}. At the coupling $u=0$,
the $\mu =0 $ quantum critical point of ${\cal L}_B$
is invariant under the transformations (\ref{xx7}), after the
replacement $\Psi_F \rightarrow \Psi_B$, and we have as before $z=2$
and\index{scaling dimension!dilute bosons}
\begin{equation}
\mbox{dim}[\Psi_B ] = d/2,\qquad
\mbox{dim}[\mu] = 2;
\label{xx40}
\end{equation}
these results will shortly be seen to be exact in all $d$.
We can  easily determine the scaling dimension of the quartic
coupling $u$ at the $u=0$, $\mu=0$ fixed point under the bosonic
analog of the transformations (\ref{xx7}); we find
\begin{equation}
\mbox{dim}[u_0] = 2-d.
\label{xx41}
\end{equation}
Thus the free-field fixed point is stable for $d>2$, in which case it is suspected
that a simple perturbative analysis of the consequences of $u$ will be adequate.
However, for $d<2$, a more careful renormalization group--based resummation
of the consequences of $u$ is required.
This identifies $d=2$ as the upper-critical dimension of the
present quantum critical point.

\begin{figure}[t]
%\epsfscale500
%\epsfxsize=225pt
\centerline{\includegraphics[width=3in]{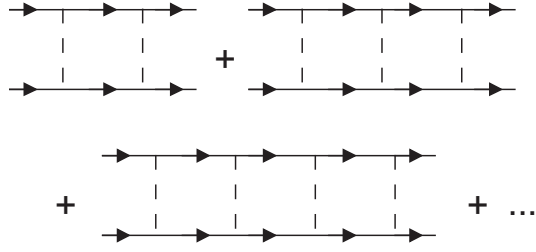}}
\caption{The ladder series of  diagrams that contribute the renormalization of the
coupling $u$ in $Z_B$ for $d< 2$.}
\label{xxf3}
\end{figure}

Our analysis of the case $d<2$ for the dilute Bose gas quantum critical
point will find, somewhat surprisingly,
that all the renormalizations,
and the associated flow equations, can be determined exactly in closed form.
We begin by considering the one-loop renormalization of the quartic coupling $u_0$
at the $\mu=0$, $T=0$ quantum critical point. It turns out that only
the ladder series of Feynman diagrams shown in Fig.~\ref{xxf3} need be considered
(the $T$ matrix\index{T matrix@$T$ matrix}).
Evaluating the first term of the series in Fig.~\ref{xxf3}
for the case of zero external frequency and momenta, we
obtain the contribution
\begin{equation}
-u_0^2 \,\int \,\frac{d \omega}{2 \pi}
\,\int \,\frac{d^d k}{(2 \pi )^d} \,\frac{1}{(- i \omega + k^2 / (2m))}
\,\frac{1}{ (i \omega + k^2 / (2m))} = -u_0^2 \,\int \,\frac{d^d k}{(2 \pi )^d}
\frac{m}{k^2}
\label{xx42}
\end{equation}
(the remaining ladder diagrams\index{ladder diagrams} are powers of (\ref{xx42}) and form
a simple geometric series).
Notice the infrared singularity for $d<2$, which is cured
by moving away from the quantum critical point, or by external momenta.

We can proceed further by a simple application of the momentum shell RG.
Note that we will apply cutoff $\Lambda$ only in momentum space. The RG
then proceeds by integrating {\em all\/} frequencies, and momentum modes
in the shell between $\Lambda e^{-\ell}$ and $\Lambda$. The renormalization of the coupling
$u_0$ is then given by the first diagram in Fig.~\ref{xxf3}, and after
absorbing some phase space factors
by a redefinition of interaction coupling
\begin{equation}
u_0 = \frac{\Lambda^{2-d}}{2 m S_d} u, \label{xx43a}
\end{equation}
we obtain~\cite{fishoh,fwgf}
\begin{equation}
\frac{du}{d \ell} = \epsilon u - \frac{u^2}{2}.
\label{xx45}
\end{equation}
Here $S_d = 2/(\Gamma(d/2) (4 \pi)^{d/2})$ is the usual phase space factor, and
\begin{equation}
\epsilon = 2 - d.
\label{xx44}
\end{equation}
Note that for $\epsilon > 0$, there is a stable fixed point at
\begin{equation}
u^{\ast} = 2 \epsilon,
\label{xx46}
\end{equation}
which will control all the universal properties of $Z_B$.

The flow equation
(\ref{xx45}),
and the fixed point value (\ref{xx46}) are {\em exact\/}
to all orders in $u$ or $\epsilon$, and it is not necessary to consider $u$-dependent
renormalizations to the field scale of $\Psi_B$ or any of the other couplings
in $Z_B$.
This result is ultimately a consequence of a very simple fact: The ground state
of $Z_B$ at the quantum critical point $\mu=0$ is simply the empty vacuum
with no particles. So any interactions that appear are entirely due to
particles that have been created by the external fields.
In particular, if we introduce the bosonic Green's function (the
analog of (\ref{xx11}))
\begin{equation}
G_B (x, t) = \big\langle \Psi_B (x,t) \Psi_B^{\dagger} (0,0) \big\rangle,
\label{xx20y}
\end{equation}
then for $\mu \leq 0$ and $T=0$, its Fourier transform $G(k, \omega)$ is
given exactly by the free field expression
\begin{equation}
G_B (k, \omega ) = \frac{1}{-\omega + k^2 / (2 m) - \mu}.
\label{xx47}
\end{equation}
The field $\Psi_B^{\dagger}$ creates a particle that travels freely until
its annihilation at $(x,t)$ by the field $\Psi_B$; there are no other
particles present at $T=0$, $\mu \leq 0$, and so the propagator
is just the free field one. The simple result (\ref{xx47}) implies that
the scaling dimensions in (\ref{xx40}) are exact. Turning to the
renormalization of $u$, it is clear from the diagram in Fig.~\ref{xxf3}
that we are considering the interactions of just two particles.
For these, the only nonzero diagrams are the one shown in Fig.~\ref{xxf3},
which involve repeated scattering of just these particles. Formally, it is possible
to write down many other diagrams that could contribute to the renormalization
of $u$; however, all of these vanish upon performing the integral over
internal frequencies for there is always one integral that can be closed
in one half of the frequency plane where the integrand has no poles.
This absence of poles is of course just a more mathematical way of stating
that there are no other particles around.

We will  consider application of these renormalization group results
separately for the cases below and above the upper-critical dimension of $d=2$.

\subsection{$d < 2$}
\label{sec:xx4}

First, let us note some important general implications of the theory controlled
by the fixed point interaction (\ref{xx46}). As we have already noted, the scaling
dimensions of $\Psi_B$ and $\mu$ are given precisely by their free field
values in (\ref{xx40}), and the dynamic exponent $z$ also retains
the tree-level value $z=2$. All these scaling dimensions are identical to those
obtained for the case of the spinless Fermi gas in Section~\ref{sec:fermigas}.
Further, the presence of a nonzero and universal interaction strength $u^{\ast}$
in (\ref{xx46}) implies
that the bosonic system is stable for the case $\mu > 0$ because
the repulsive interactions
will prevent the condensation of infinite density of bosons (no such interaction
was necessary for the fermion case, as the Pauli exclusion was already
sufficient to stabilize the system).
These two facts imply that the formal scaling structure of the
bosonic fixed point being considered here is identical to that of the
fermionic one considered in Section~\ref{sec:fermigas} and that the scaling forms
of the two theories are {\em identical}. In particular, $G_B$ will obey
a scaling form identical to that for $G_F$ in (\ref{xx10}) (with a
corresponding scaling function
$\Phi_{G_B}$),
while the free energy, and associated
derivatives, obey (\ref{xx10a}) (with a scaling function $\Phi_{{\cal F}_B}$).
The
universal functions $\Phi_{G_B}$ and $\Phi_{{\cal F}_B}$ can be
determined order by order in the present $\epsilon = 2-d$ expansion, and this
will be illustrated shortly.

Although the fermionic and bosonic fixed points share the same scaling dimensions,
they are distinct fixed points for general $d < 2$. However, these two fixed points are identical precisely in
$d=1$~\cite{sss}.
Evidence for this was presented in Ref.~\cite{dsbose}, where
the anomalous dimension of the composite operator $\Psi_B^2$ was computed exactly
in the $\epsilon$ expansion and was found to be identical to that of the
corresponding fermionic operator. Assuming the identity of the fixed points,
we can then make a stronger statement about the universal scaling function:
those for the free energy (and all its derivatives) are identical
$\Phi_{{\cal F}_B} = \Phi_{{\cal F}_F}$ in $d=1$.
In particular, from (\ref{xx12}) and (\ref{xx12z}) we conclude that the
boson density is given by
\begin{equation}
\langle Q \rangle = \big\langle \Psi_B^{\dagger} \Psi_B \big\rangle =
\int \frac{dk}{2 \pi} \frac{1}{e^{(k^2/(2m) - \mu)/T} + 1}
\label{xx47a}
\end{equation}
in $d=1$ only.
The operators $\Psi_B$ and
$\Psi_F$ are still distinct and so there is no reason for the scaling functions
of their correlators to be the same. However, in $d=1$, we can relate the 
universal scaling function of $\Psi_B$ to those of $\Psi_F$ via
a continuum version of the Jordan-Wigner transformation
\begin{equation}
\Psi_B (x, t) = \exp \left( i \pi \int_{-\infty}^x dy \Psi_F^{\dagger} (y,t)
\Psi_F (y,t) \right) \Psi_F (x,t).
\label{xx20z}
\end{equation}
This identity is applied to obtain numerous exact results in Ref.~\cite{book}

As not all observables can be computed exactly in $d=1$ by the mapping to the
free fermions, we will now consider the
$\epsilon=2-d$ expansion. We will present a simple $\epsilon$ expansion %
calculation~\cite{senthil}
for illustrative purposes. We focus on density of bosons at $T=0$. Knowing that
the free energy obeys the analog of (\ref{xx10a}), we can conclude that a relationship
like (\ref{xx5}) holds:
\begin{equation}
\big\langle \Psi^{\dagger}_B \Psi_B \big\rangle =  \left\{
\begin{array}{c@{\quad}c}
{\cal C}_d (2 m \mu)^{d/2}, & \mu > 0, \\[3pt]
0, & \mu < 0,
\end{array}\right.
\label{xx48}
\end{equation}
at $T=0$, with ${\cal C}_d$ a universal number. The identity of the bosonic
and fermionic theories in $d=1$
implies from (\ref{xx5}) or from (\ref{xx47a})
that ${\cal C}_1 = S_1/1 = 1/\pi$.
We will  show how to compute ${\cal C}_d$ in the $\epsilon$ expansion; similar
techniques can be used for almost any observable.

Even though the position of the fixed point is known exactly in
(\ref{xx46}), not all observables can be computed exactly 
because they have
contributions to arbitrary order in $u$. 
However, universal results can be obtained order-by-order in $u$,
which then become a power series in $\epsilon = 2-d$. As an example,
let us examine the low order contributions to the boson density.
To compute the boson density for $\mu > 0$, we anticipate that there is
condensate of the boson field $\Psi_B$, and so we write
\begin{equation}
\Psi_B (x,\tau) = \Psi_0 + \Psi_1 (x,t),
\label{xx49}
\end{equation}
where $\Psi_1$ has no zero wavevector and frequency component.
Inserting this into ${\cal L}_B$ in (\ref{xx0}), and expanding to second order
in $\Psi_1$, we get
\begin{eqnarray}
 {\cal L}_1 &=& - \mu |\Psi_0|^2 + \frac{u_0}{2} |\Psi_0|^4
- \Psi_1^{\ast} \frac{\partial \Psi_1}{\partial \tau}
 + \frac{1}{2m} \left| \nabla \Psi_1 \right|^2 \nonumber \\
&&-\mu |\Psi_1|^2 + \frac{u_0}{2}
\left( 4 |\Psi_0|^2 |\Psi_1|^2 + \Psi_0^2 \Psi_1^{\ast 2}
+ \Psi_0^{\ast 2} \Psi_1^2 \right).
\label{xx50}
\end{eqnarray}
This is a simple quadratic theory in the canonical Bose field $\Psi_1$,
and its spectrum and ground state energy
can be determined by the familiar Bogoliubov transformation.
Carrying out this step, we obtain the following formal expression for the
free energy density ${\cal F}$ as a function of the condensate $\Psi_0$
at $T=0$:
\begin{eqnarray}
 {\cal F} ( \Psi_0 ) &=&  - \mu |\Psi_0|^2 + \frac{u_0}{2} |\Psi_0|^4
+ \frac{1}{2}
\int \frac{d^d k}{( 2\pi )^d}
\Biggl[ \left\{ \left(\frac{k^2}{2m} - \mu + 2 u_0 |\Psi_0|^2 \right)^2\right.
-\left.\vphantom{L^{L^{L^{L^{L^{L^L}}}}}} u_0^2 |\Psi_0|^4 \right\}^{1/2}\nonumber\\
&& - \left(\frac{k^2}{2m} - \mu + 2 u_0 |\Psi_0|^2 \right)
\Biggr].\label{xx51}
\end{eqnarray}
To obtain the physical free energy density, we have to minimize ${\cal F}$ with respect
to variations in $\Psi_0$ and to substitute the result back into (\ref{xx51}).
Finally, we can take the derivative of the resulting expression with respect to $\mu$
and obtain the required expression for the boson density, correct to the first two
orders in $u_0$:
\begin{equation}
\big\langle \Psi_B^{\dagger} \Psi_B \big\rangle =
\frac{\mu}{u_0} + \frac{1}{2} \int \frac{d^d k}{(2 \pi)^d}
\left[ 1 - \frac{k^2}{\sqrt{k^2 ( k^2 + 4 m \mu )}} \right].
\label{xx52}
\end{equation}

To convert (\ref{xx52}) into a universal result, we need to evaluate it at the coupling
appropriate to the fixed point (\ref{xx46}). This is most easily done by the 
field-theoretic RG. So let us translate the RG
equation (\ref{xx45}) into this language. 
We introduce a momentum scale $\tilde{\mu}$ (the tilde is to prevent
confusion with the chemical potential) and express $u_0$ in terms of a
dimensionless coupling $u_R$ by
\begin{equation}
u_0= u_R \frac{(2m) \tilde{\mu}^{\epsilon}}{S_d}
\left( 1 + \frac{u_R}{2 \epsilon} \right).
\label{xx43}
\end{equation}
The motivation behind the choice of the renormalization factor in (\ref{xx43})
is that the renormalized four-point coupling,
when expressed in terms of $u_R$, and evaluated in $d=2-\epsilon$,
is free of poles in $\epsilon$ as can easily be explicitly checked using
(\ref{xx42}) and the associated geometric series.
Then, we evaluate (\ref{xx52}) at the fixed point value of $u_R$, 
compute any physical observable as a formal
diagrammatic expansion in $u_0$, substitute $u_0$ in favor of $u_R$
using (\ref{xx43}), and expand the resulting expression
in powers of $\epsilon$. All poles in $\epsilon$ should cancel,
but the resulting expression will depend upon the arbitrary
momentum scale $\tilde{\mu}$. At the fixed point value
$u_R^{\ast}$, dependence upon $\tilde{\mu}$
then disappears and
a universal answer remains.
In this manner we obtain from (\ref{xx52}) a 
universal expression in the form (\ref{xx48}) with
\begin{equation}
{\cal C}_d = S_d
\left[\frac{1}{2 \epsilon} + \frac{\ln 2 - 1}{4} + {\cal O} ( \epsilon )\right].
\label{xx53}
\end{equation}

\subsection{$d=3$}
\label{sec:xxd3}

Now we briefly discuss $2 < d < 4$: details appear elsewhere \cite{book}.
In $d=2$, the upper critical dimension, there are logarithmic corrections which
were computed by Prokof'ev {\em et al.} \cite{prokofev}.
Related results, obtained through somewhat different methods, are available in
the literature~\cite{popov1,popov2,fishoh,sss}.

\begin{figure}[t]
%\epsfscale360
%\epsfxsize=140pt
\centerline{\includegraphics[width=3.6in]{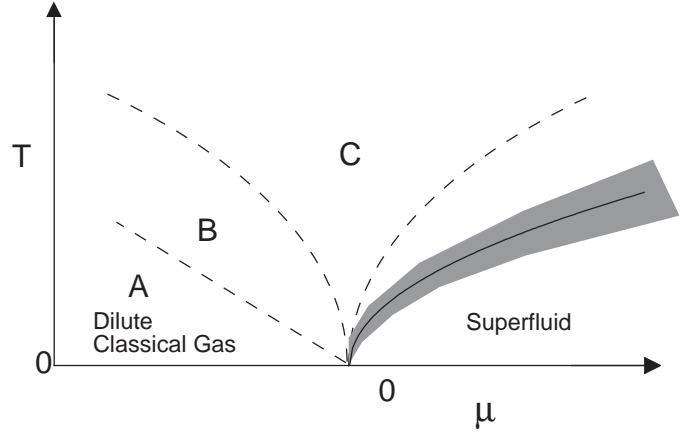}}
\caption{Crossovers of the dilute Bose gas in $d=3$ as a function of the
chemical potential $\mu$ and the temperature $T$. The regimes labeled A, B, C
are described in Ref.~\cite{book}. The solid line is the finite-temperature phase transition
where the superfluid order disappears; the shaded region is where there
is an effective classical description of thermal fluctuations. The contours of
constant density are similar to those in Fig.~\protect\ref{xxf2} and are not displayed. }
\label{xxf4}
\end{figure}

The quantum critical point at $\mu =0$, $T=0$
is above its upper-critical dimension, and we expect mean-field theory to apply.
The analog of the mean-field result in the present context is the $T=0$ relation
for the density
\begin{equation}
\big\langle \Psi^{\dagger}_B \Psi_B \big\rangle =  \left\{
\begin{array}{c@{\quad}c}
\mu / u_0 + \cdots, & \mu > 0, \\[2pt]
0, & \mu < 0,
\end{array}
\right. 
\label{xx54}
\end{equation}
where the ellipses represents terms that vanish faster as $\mu \rightarrow 0$.
Notice that this expression for the density is not universally dependent upon
$\mu$; rather it depends upon the strength of the two-body interaction $u_0$ (more precisely,
it can be related to the $s$-wave scattering length
$a$ by $u_0 = 4 \pi a/m$).\index{scattering length}
The crossovers and phase transitions at $T>0$ are
sketched in Fig.~\ref{xxf4}. 
These are similar to those of the spinless Fermi gas, but 
now there can be a phase transition within one of the regions.
Explicit expressions for the crossovers \cite{book} have been 
presented by Rasolt et al.~\cite{rasolt}, Weichman et al.~\cite{rasolt2}
and also addressed in earlier work~\cite{kk1,kk2,creswick}.

\section{The Dilute Spinful Fermi Gas: the Feshbach Resonance}
\label{sec:feshbach}

This section turns to the case of the spinful Fermi gas with short-range interactions;
as we noted in the introduction, this is a problem which has acquired renewed importance
because of the new experiments on ultracold fermionic atoms. 

The partition function of the theory examined in this section was displayed in 
(\ref{fesh1}). 
The renormalization group properties of this theory in the zero density limit are {\em identical\/}
to those the dilute Bose gas considered in Section~\ref{sec:xx3}. The scaling dimensions of the couplings
are the same, the scaling dimension of $\Psi_{F\sigma}$ is $d/2$ as for $\Psi_B$ in (\ref{xx40}), 
and the flow of the $u$ is given by (\ref{xx45}). Thus for $d<2$, a spinful Fermi gas
with repulsive interactions is described by the stable fixed point in (\ref{xx46}).

\begin{figure}
\centerline{\includegraphics[width=180pt]{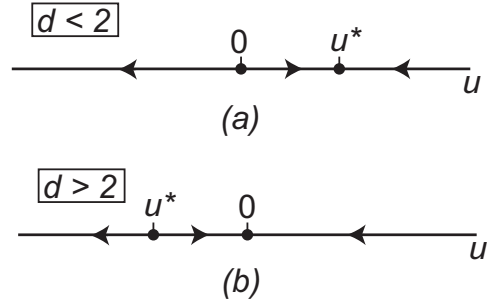}}
\caption{The exact RG flow of (\ref{xx45}).  ({\em a\/}) For $d<2$ ($\epsilon>0$), the infrared stable fixed point
  at $u=u^\ast > 0$ describes quantum liquids of either bosons or fermions
  with repulsive interactions which are
  generically universal in the low density limit. In $d=1$ this fixed point is described
  by the spinless free Fermi gas (`Tonks' gas), for all statistics and spin of the
  constituent particles. ({\em b\/}) For $d>2$ ($\epsilon<0$) the
  infrared unstable fixed point at $u=u^\ast < 0$ describes the Feshbach
  resonance which obtains for the case of attractive interactions. The relevant perturbation
  $(u-u^\ast)$ corresponds to the the detuning from the resonant
  interaction.}
\label{fig:feshbach}
\end{figure}

However, for the case of spinful Fermi gas case, we can consider another regime of parameters
which is of great experimental importance. We can also allow $u$ to be attractive: unlike the Bose
gas case, the $u<0$ case is not immediately unstable, because the Pauli exclusion principle can stabilize
a Fermi gas even with attractive interactions. Furthermore, at the same time we should also consider
the physically important case with $d>2$, when $\epsilon < 0$. The distinct
nature of the RG flows predicted by (\ref{xx45}) for the two signs of $\epsilon$ are shown in 
Fig.~\ref{fig:feshbach}.

Notice the {\em unstable\/} fixed point present for $d>2$ and $u<0$. Thus accessing the fixed point requires
fine-tuning of the microscopic couplings. As discussed in Refs.~\cite{nishida,predrag}, this fixed point
describes a Fermi gas at a {\em Feshbach resonance,\/} where the interaction between the fermions is universal.
For $u<u^\ast$, the flow is to $u \rightarrow -\infty$: this corresponds to a strong attractive interaction between the fermions,
which then bind into tightly bound pairs of bosons, which then Bose condense; this corresponds to the
so-called `BEC' regime.
On the other hand, for $u > u^\ast$, the flow is to $u \nearrow 0$, and the weakly interacting fermions then form
the Bardeen-Cooper-Schrieffer (BCS) superconducting state. 

Note that the fixed point at $u=u^\ast$ for $Z_{Fs}$ has {\em two} relevant directions for $d>2$.
As in the other problems considered earlier, one corresponds to the chemical potential $\mu$. The other corresponds
to the deviation from the critical point $u - u^\ast$, and this (from (\ref{xx45})) has RG eigenvalue 
$-\epsilon = d-2 > 0$. This perturbation corresponds to the ``detuning'' from the Feshbach resonance, $\nu$
(not to be confused with the symbol for the correlation length exponent); we have $\nu \propto u - u^\ast$.
Thus we have
\begin{equation}
\mbox{dim}[\mu] = 2~~,~~\mbox{dim}[\nu] = d-2.
\label{fesh0}
\end{equation}
 These two relevant perturbations
will have important consequences for the phase diagram, as we will see shortly.

For now, let us understand the physics of the Feshbach resonance better. For this, it is useful to compute the two body
$T$ matrix exactly by summing the graphs in Fig.~\ref{xxf3}, along with a direct interaction first order in $u_0$.
The second order term was already evaluated for the bosonic case in (\ref{xx42}) for zero external momentum and frequency, 
and has an identical value for the
present fermionic case. Here, however, we want the off-shell $T$-matrix, for the case in which the incoming particles 
have momenta $k_{1,2}$, and frequencies $\omega_{1,2}$. Actual for the simple momentum-independent interaction $u_0$, the $T$ matrix depends only upon the sums $k=k_1+ k_2$ and $\omega=\omega_1 + \omega_2$, and is independent
of the final state of the particles, and the diagrams in Fig.~\ref{xxf3} form a geometric series. In this manner we obtain
\begin{eqnarray}
&& \frac{1}{T (k, i \omega )}  = \frac{1}{u_0}  \nonumber \\
&&~~+ \int \,\frac{d \Omega}{2 \pi}
\int \frac{d^d p}{(2 \pi )^d} \,\frac{1}{(- i (\Omega+\omega) + (p+k)^2 / (2m))}
\,\frac{1}{ (i \Omega + p^2 / (2m))} \nonumber \\
&&= \frac{1}{u_0} +  \int_0^\Lambda \frac{d^d p}{(2 \pi )^d} \frac{m}{p^2}+ \frac{\Gamma (1-d/2)}{(4 \pi)^{d/2}}
m^{d/2} \left[ - i \omega + \frac{k^2}{4m} \right]^{d/2-1}. \label{fesh2}
\end{eqnarray}
In $d=3$, the $s$-wave scattering amplitude of the two particles, $f_0$, is related to the $T$-matrix at zero center
of mass momentum and frequency $k^2/m$ by $f_0 (k) = - m T(0, k^2 /m)/(4 \pi)$, and so we obtain
\begin{equation}
f_0 (k) = \frac{1}{-1/a - ik} \label{fesh3}
\end{equation}
where the scattering length, $a$, is given by
\begin{equation}
\frac{1}{a} = \frac{4 \pi}{m u_0} + \int_0^\Lambda \frac{d^3 p}{(2 \pi )^3} \frac{4 \pi}{p^2}. \label{fesh4}
\end{equation}
For $u_0 < 0$, we see from (\ref{fesh4}) that there is a critical value of $u_0$ where the scattering length
diverges and changes sign: this is the Feshbach resonance. 
We identify this critical value with the fixed point $u=u^\ast$ of the RG flow (\ref{xx45}). It is conventional
to identify the deviation from the Feshbach resonance by the detuning $\nu$ 
\begin{equation}
\nu \equiv - \frac{1}{a}.
\label{fesh5}
\end{equation}
Note that $\nu \propto u - u^\ast$, as claimed earlier. 
For $\nu > 0$, we have weak attractive interactions, and the scattering length is negative.
For $\nu < 0$, we have strong attractive interactions, and a positive scattering length.
Importantly, for $\nu < 0$, there is a two-particle bound state, whose energy can be deduced from the
pole of the scattering amplitude; recalling that the reduced mass in the center of mass frame is $m/2$,
we obtain the bound state energy, $E_b$
\begin{equation}
E_b = - \frac{\nu^2}{m}. \label{fesh6}
\end{equation}

\begin{figure}
\centerline{\includegraphics[width=180pt]{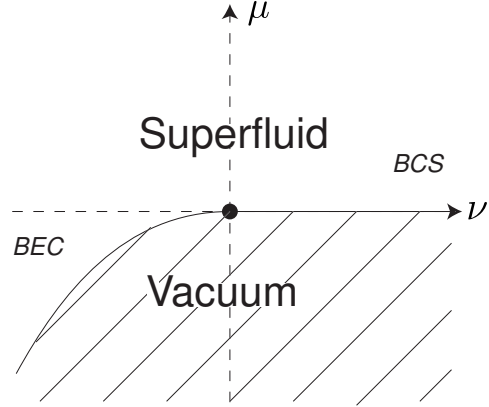}}
\caption{Universal phase diagram at zero temperature
for the spinful Fermi
gas in $d=3$ as a function of the chemical potential $\mu$
and the detuning $\nu$. The vacuum state (shown hatched) has no particles. The position of the
$\nu < 0 $ phase boundary is determined by the energy of the
two-fermion bound state in (\ref{fesh6}): $\mu = - \nu^2/(2m)$. The density of particles vanishes
continuously at the second order quantum phase transition boundary
of the superfluid phase, which is indicated by the thin continuous
line. The quantum multicritical point at $\mu=\nu=0$ (denoted by
the filled circle) controls all the universal physics of the dilute spinful Fermi gas near
a Feshbach resonance. The universal properties of the critical line $\mu=0$, $\nu > 0$
map onto the theory of Section~\ref{sec:fermigas}, while those of the critical line
$\mu = - \nu^2 / (2m)$, $\nu < 0$ map onto the theory of Section~\ref{sec:xx3}. This implies that the $T>0$
crossovers in Fig.~\ref{xxf2} apply for $\nu > 0$ (the ``Fermi liquid'' region of Fig.~\ref{xxf2} now has BCS superconductivity
at an exponentially small $T$), while those of Fig.~\ref{xxf4} apply for $\nu < 0$.}
\label{fig:unitary}
\end{figure}

We can now draw the zero temperature  
phase diagram \cite{predrag} of $Z_{Fs}$ as a function of $\mu$ and $\nu$, and the result is shown in 
 Fig.~\ref{fig:unitary}.

For $\nu > 0$, there is no bound state, and so no fermions are present for $\mu < 0$.
At $\mu =0$, we have an onset of non-zero fermion density, just as in the other sections. 
These fermions experience a weak attractive interaction, and so 
experience the Cooper instability once there is a finite density of fermions for $\mu > 0$.
So the ground state for $\mu > 0$ is a paired Bardeen-Cooper-Schrieffer (BCS) superfluid,
as indicated in Fig.~\ref{fig:unitary}. For small negative scattering lengths, the BCS state modifies the fermion state only near the Fermi level. Consequently as $\mu \searrow 0$ (specifically for $\mu < \nu^2/m$), we can 
neglect the pairing in computing the fermion density. We therefore conclude that the
universal critical properties of the line $\mu=0$, $\nu > 0$ map precisely on to two copies (for the spin 
degeneracy) of the non-interacting
fermion model $Z_F$ studied in Section~\ref{sec:fermigas}. In particular the $T>0$ properties for $\nu>0$
will map onto the crossovers in Fig.~\ref{xxf2}. The only change is that the BCS pairing instability will appear
below an exponentially small $T$ in the ``Fermi liquid'' regime. However, the scaling functions for the density as
a function of $\mu/T$ will remain unchanged.

For $\nu < 0$, the situation changes dramatically. Because of the presence of the bound state
(\ref{fesh6}), it will pay to introduce fermions even for $\mu < 0$. The chemical potential for a fermion
pair is $2 \mu$, and so the threshold for having a non-zero density of paired fermions is $\mu = E_b/2$.
This leads to the phase boundary shown in Fig.~\ref{fig:unitary} at $\mu = - \nu^2 / (2m)$. 
Just above the phase boundary, the density of fermion pairs in small, and so these can be treated
as canonical bosons. Computations of the interactions between these bosons \cite{predrag} show that
they are repulsive. Therefore we map their dynamics onto those of the dilute Bose gas
studied in Section~\ref{sec:xx3}. Thus the universal properties of the critical line $\mu = - \nu^2 / (2m)$
are equivalent to those of $Z_B$. Specifically, this means that the $T>0$ properties across this critical line
map onto those of Fig.~\ref{xxf4}.

Thus we reach the interesting conclusion that the Feshbach resonance at $\mu=\nu=0$ is a multicritical
point separating the density onset transitions of $Z_F$ (Section~\ref{sec:fermigas}) and $Z_B$ (Section~\ref{sec:xx3}).
This conclusion can be used to sketch the $T>0$ extension of Fig.~\ref{fig:unitary}, on either side of the
$\nu=0$ line.

We now need a practical method of computing universal properties of $Z_{Fs}$ near the $\mu = \nu=0$ fixed point, including
its crossovers into the regimes described by $Z_F$ and $Z_B$. The fixed point (\ref{xx45}) of $Z_{Fs}$ provides
an expansion of the critical theory in the powers of $\epsilon = 2-d$. However, observe from Fig.~\ref{fig:feshbach},
the flow for $u < u^\ast$ is to $u \rightarrow -\infty$. The latter flow describes the crossover into the dilute
Bose gas theory, $Z_B$, and so this cannot be controlled by the $2-d$ expansion. The following subsections
will propose two alternative analyses of the Feshbach resonant fixed point which will address this difficulty.

\subsection{The Fermi-Bose Model}
\label{sec:fb}

One successful approach is to promote the two fermion bound state in (\ref{fesh6}) to a canonical boson field $\Psi_B$. 
This boson should also be able to mix with the scattering states of two fermions. We are therefore led 
to consider the following model
\begin{eqnarray}
Z_{FB} &=& \,\int {\cal D} \Psi_{F\uparrow} (x,\tau) {\cal D} \Psi_{F\downarrow} (x,\tau) {\cal D} \Psi_{B} (x, \tau)
\exp\left( - \,\int d \tau d^d x \, {\cal L}_{FB} \right),
 \nonumber \\
{\cal L}_{FB} &=&  \Psi_{F\sigma}^{\ast} \,\frac{\partial \Psi_{F\sigma}}{\partial \tau}
 + \,\frac{1}{2m} \left| \nabla \Psi_{F\sigma} \right|^2 -\mu |\Psi_{F\sigma}|^2 \nonumber \\
 &+& \Psi_{B}^{\ast} \,\frac{\partial \Psi_{B}}{\partial \tau}
 + \,\frac{1}{4m} \left| \nabla \Psi_{F\sigma} \right|^2 + (\delta -2\mu) |\Psi_{B}|^2 \nonumber \\
 &-& \lambda_0 \left( \Psi_B^\ast \Psi_{F \uparrow} \Psi_{F \downarrow} + \Psi_B \Psi_{F \downarrow}^\ast 
 \Psi_{F \uparrow}^\ast \right).
\label{fesh7}
\end{eqnarray}
Here we have taken the bosons to have mass $2m$, because that is the expected mass of the two-fermion bound
state by Galilean invariance. We have omitted numerous possible quartic terms between
the bosons and fermions above, and these will turn out to be irrelevant in the analysis below.

The conserved U(1) charge for $Z_{FB}$ is
\begin{equation}
Q = \Psi_{F \uparrow}^\ast \Psi_{F \uparrow} + \Psi_{F \downarrow}^\ast \Psi_{F \downarrow} + 
2 \Psi_{B}^\ast \Psi_{B} ,
\label{fesh8}
\end{equation}
and so $Z_{FB}$ is in the class of models being studied here.
The factor of 2 in (\ref{fesh8}) accounts for the $2 \mu$ chemical potential for the bosons in 
(\ref{fesh7}). 
For $\mu$ sufficiently negative it is clear that $Z_{FB}$ will have neither fermions nor bosons present,
and so $\langle Q \rangle = 0$. Conversely for positive $\mu$, we expect $\langle Q \rangle \neq 0$,
indicating a transition as a function of increasing $\mu$. Furthermore, for $\delta$ large and positive, the
$Q$ density will be primarily fermions, while for $\delta $ negative the $Q$ density will be mainly bosons;
thus we expect a Feshbach resonance at intermediate values of $\delta$, which then plays
the role of detuning parameter.

We have thus argued that the phase diagram of $Z_{FB}$ as a function of $\mu$ and $\delta$
is qualitatively similar to that
in Fig.~\ref{fig:unitary}, with a Feshbach resonant multicritical point near the center. The main claim
of this section is that the universal properties of $Z_{FB}$ and $Z_{Fs}$ are {\em identical\/} near this
multicritical point \cite{predrag,nishida}. Thus, in a strong sense, the theories $Z_{FB}$ and $Z_{Fs}$ are equivalent.
Unlike the equivalence between $Z_B$ and $Z_F$, which held only in $d=1$, the present 
equivalence applies for $d > 2$.

We will establish the equivalence by an exact RG analysis of the zero density critical theory.
We scale the spacetime co-ordinates and the fermion field as in (\ref{xx7}), but allow
an anomalous dimension $\eta_b$ for the boson field relative to (\ref{xx40}):
\begin{eqnarray}
x' &=& x e^{-\ell}, \nonumber \\
\tau' &=& \tau e^{-z\ell}, \nonumber \\
\Psi'_{F\sigma} &=& \Psi_{F \sigma} e^{d \ell/2}, \nonumber \\
\Psi'_{B} &=& \Psi_{B} e^{(d + \eta_b) \ell/2} \nonumber \\
\lambda'_0 &=& \lambda_0 e^{(4 - d - \eta_b) \ell/2}
\label{fesh9}
\end{eqnarray}
where, as before, we have $z=2$.
At tree level, the theory $Z_{FB}$ with $\mu = \delta=0$ is invariant under the transformations
in (\ref{fesh9}) with $\eta_b = 0$. At this level, we see that the coupling $\lambda_0$
is relevant for $d<4$, and so we will have to consider the influence of $\lambda_0$.
This also suggests that we may be able
to obtain a controlled expansion in powers of $(4-d)$.

\begin{figure}
\centerline{\includegraphics[width=2.5in]{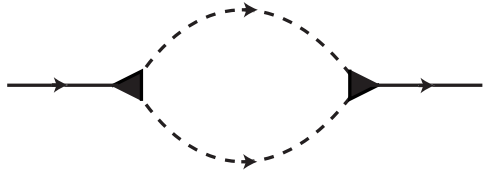}}
\caption{Feynman diagram contributing to the RG. 
The dark triangle is the $\lambda_0$ vertex, the full line is the $\Psi_B$
propagator, and the dashed line is the $\Psi_F$ propagator.} \label{fig:fesh1}
\end{figure}

Upon considering corrections in powers of $\lambda_0$
in the critical theory, it is not difficult to show that there is a non-trivial contribution 
from only a single Feynman diagram: this is the self -energy diagram for $\Psi_B$
which is shown in Fig.~\ref{fig:fesh1}. 
All other diagrams vanish in the zero density theory, for reasons similar
to those discussed for $Z_B$ below (\ref{xx46}). This diagram is closely related to the integrals in the $T$-matrix
computation in (\ref{fesh2}), and leads to the following contribution to the boson self energy $\Sigma_B$:
\begin{eqnarray}
&& \Sigma_B (k, i \omega) \nonumber \\
&& =  \lambda_0^2 \int \frac{d \Omega}{2 \pi}
\int_{\Lambda e^{-\ell}}^{\Lambda} 
\frac{d^d p}{(2 \pi )^d} \frac{1}{(- i (\Omega+\omega) + (p+k)^2 / (2m))}
\,\frac{1}{ (i \Omega + p^2 / (2m))} \nonumber \\
&& =  \lambda_0^2 
\int_{\Lambda e^{-\ell}}^{\Lambda} 
\frac{d^d p}{(2 \pi )^d} \frac{1}{(- i \omega + (p+k)^2 / (2m) + p^2 / (2m))} \nonumber \\
&& =  \lambda_0^2 
\int_{\Lambda e^{-\ell}}^{\Lambda} 
\frac{d^d p}{(2 \pi )^d} \frac{m}{p^2} - \lambda_0^2 \left( - i \omega  + \frac{k^2}{4m} \left( 2 - \frac{4}{d} \right) \right)
\int_{\Lambda e^{-\ell}}^{\Lambda} 
\frac{d^d p}{(2 \pi )^d} \frac{m^2}{p^4}. \label{fesh10}
\end{eqnarray}
The first term is a constant that can absorbed into a redefinition of $\delta$. For the first time, 
we see above a special role for the spatial dimension $d=4$, where the momentum integral is logarithmic.
Our computations below will turn to be an expansion in powers of $(4-d)$, and so we will evaluate
the numerical prefactors in (\ref{fesh10}) with $d=4$. The result turns out to be correct
to all orders in $(4-d)$, but to see this explicitly we need to use a proper Galilean-invariant
cutoff in a field theoretic approach \cite{predrag}. The simple momentum shell method
being used here preserves Galilean invariance only in $d=4$.

With the above reasoning, we see that the second term in the boson self-energy in (\ref{fesh10}) can be
absorbed into a rescaling of the boson field under the RG. We therefore find a non-zero anomalous
dimension
\begin{equation}
\eta_b =  \lambda^2 , \label{fesh11}
\end{equation}
where we have absorbed phase space factors into the coupling $\lambda$ by
\begin{equation}
\lambda_0 = \frac{\Lambda^{2-d/2}}{m \sqrt{S_d}} \lambda . \label{fesh12}
\end{equation}

With this anomalous dimension, we use (\ref{fesh9}) to obtain the exact RG equation
for $\lambda$:
\begin{equation}
\frac{d \lambda}{d \ell} = \frac{(4-d)}{2} \lambda - \frac{\lambda^3}{2} . \label{fesh13}
\end{equation}
For $d<4$, this flow has a stable fixed point at $\lambda = \lambda^\ast = \sqrt{(4-d)}$.
The central claim of this subsection is that the theory $Z_{FB}$ at this fixed point is identical
to the theory $Z_{Fs}$ at the fixed point $u=u^\ast$ for $2 < d < 4$.

Before we establish this claim, note that at the fixed point, we obtain the exact result for
the anomalous dimension of the the boson field
\begin{equation}
\eta_b = 4-d .\label{fesh14}
\end{equation}

\begin{figure}
\centerline{\includegraphics[width=2in]{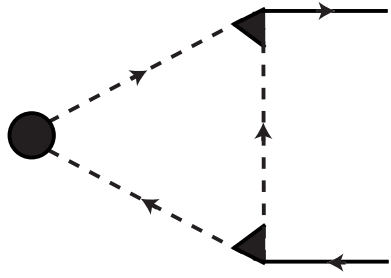}}
\caption{Feynman diagram for the mixing between the renormalization of the $\Psi_F^\dagger \Psi_F$
and $\Psi_B^\dagger \Psi_B$ operators. The filled circle is the $\Psi_F^\dagger \Psi_F$ source.
Other notation is as in Fig.~\ref{fig:fesh1}.} \label{fig:fesh2}
\end{figure}

Let us now consider the spectrum of relevant perturbations to the $\lambda = \lambda^\ast$ fixed
point. As befits a Feshbach resonant fixed point, there are 2 relevant perturbations in $Z_{FB}$, the detuning 
parameter $\delta$
and the chemical potential $\mu$. Apart from the tree level rescalings, at one loop
we have the diagram shown in Fig.~\ref{fig:fesh2}. 
This diagram has a $\Psi_{F \sigma}^{\dagger} \Psi_{F \sigma}$
source, and it renormalizes the co-efficient of $\Phi^{\dagger} \Phi$; it evaluates to
\begin{eqnarray}
&& 2 \lambda_0^2 \int \frac{d \Omega}{2 \pi} \int_{\Lambda e^{-\ell}}^{\Lambda} \frac{d^d p}{(2 \pi)^d} \frac{1}{(- i \Omega + p^2 /(2m))^2 (i \Omega + p^2 /(2m))} 
\nonumber \\
&&~~~~= 2 \lambda_0^2 \int_{\Lambda e^{-\ell}}^{\Lambda} \frac{d^d p}{(2 \pi)^d} \frac{m^2}{p^4} . \label{fesh15}
\end{eqnarray}
Combining (\ref{fesh15}) with the tree-level rescalings, we obtain the RG flow equations
\begin{eqnarray}
\frac{d \mu}{d \ell} &=& 2 \mu \nonumber \\
\frac{d}{d \ell} (\delta - 2 \mu ) &=& (2 - \eta_b) (\delta - 2 \mu) - 2 \lambda^2 \mu, \label{fesh16}
\end{eqnarray} 
where the last term arises from (\ref{fesh15}).
With the value of $\eta_b$ in (\ref{fesh11}), the second equation simplifies to
\begin{equation} 
\frac{d \delta}{d \ell} = (2 - \lambda^2) \delta. \label{fesh17}
\end{equation}
Thus we see that $\mu$ and $\delta$ are actually eigen-perturbations of the fixed point at $\lambda = \lambda^\ast$,
and their scaling dimensions are
\begin{equation}
\mbox{dim}[\mu] = 2~~,~~\mbox{dim}[\delta] = d-2.
\label{fesh18}
\end{equation}
Note that these eigenvalues coincide with those of $Z_{Fs}$ in (\ref{fesh0}), with $\delta$ identified as
proportional to the detuning $\nu$. This, along with the symmetries
of $Q$ conservation and Galilean invariance, establishes the equivalence of the fixed points of $Z_{FB}$
and $Z_{Fs}$. 

The utility of the present $Z_{FB}$ formulation is that it can provide a description of universal  properties of the
unitary Fermi gas in $d=3$ via an expansion in $(4-d)$. Further details of explicit computations can 
be found in Ref.~\cite{nishida}.

\subsection{Large $N$ expansion}
\label{sec:predrag}

We now return to the model $Z_{Fs}$ in (\ref{fesh1}), and examine it in the limit of a large
number of spin components \cite{predrag,veillette}. We also use the structure of the large $N$ perturbation theory
to obtain exact results relating different experimental observable of the unitary Fermi gas.

The basic idea of the large $N$ expansion is to endow the fermion with an additional flavor index $a = 1 \ldots N/2$,
to the fermion field is $\Psi_{F \sigma a}$, where we continue to have $\sigma=\uparrow, \downarrow$.
Then, we write $Z_{Fs}$ as
\begin{equation}
\begin{array}{rcl}
Z_{Fs} &=& \,\int {\cal D}  \Psi_{F\sigma a} (x,\tau)
\exp\left( - \,\int_0^{1/T} d \tau \,\int d^d x \, {\cal L}_{Fs} \right),
 \\[16pt]
{\cal L}_{Fs} &=&  \Psi_{F\sigma a}^{\ast} \,\frac{\partial \Psi_{F\sigma a}}{\partial \tau}
 + \,\frac{1}{2m} \left| \nabla \Psi_{F\sigma a} \right|^2 -\mu |\Psi_{F\sigma a}|^2  \\[16pt]
&~&~~~~~ + \displaystyle \frac{2 u_0}{N} \Psi_{F\uparrow a}^\ast
 \Psi_{F \downarrow a}^\ast \Psi_{F\downarrow b} \Psi_{F \uparrow b}.
\end{array}
\label{fesh19}
\end{equation}
where there is implied sum over $a,b = 1 \ldots N/2$. The case of interest has $N=2$, but we will consider the limit
of large even $N$, where the problem becomes tractable.

As written, there is an evident O($N/2$) symmetry in $Z_{Fs}$ corresponding to rotations in flavor space.
In addition, there is U(1) symmetry associated with $Q$ conservation, and a SU(2) spin rotation symmetry.
Actually, the spin and flavor symmetry combine to make the global symmetry U(1)$\times$Sp($N$), but we will
not make much use of this interesting observation.

The large $N$ expansion proceeds by decoupling the quartic term in 
(\ref{fesh19}) by a Hubbard-Stratanovich transformation. 
For this we introduce a complex bosonic field $\Psi_B (x, \tau)$ and write
\begin{equation}
\begin{array}{rcl}
Z_{Fs} &=& \,\int {\cal D}  \Psi_{F\sigma a} (x,\tau) {\cal D} \Psi_B (x, \tau)
\exp\left( - \,\int_0^{1/T} d \tau \,\int d^d x \, \widetilde{\cal L}_{Fs} \right),
 \\[16pt]
\widetilde{\cal L}_{Fs} &=&  \Psi_{F\sigma a}^{\ast} \,\frac{\partial \Psi_{F\sigma a}}{\partial \tau}
 + \,\frac{1}{2m} \left| \nabla \Psi_{F\sigma a} \right|^2 -\mu |\Psi_{F\sigma a}|^2  \\[16pt]
&~& +  \displaystyle \frac{N}{2 |u_0|} |\Psi_B|^2  - \Psi_B \Psi_{F\uparrow a}^\ast
 \Psi_{F \downarrow a}^\ast  - \Psi_B^\ast \Psi_{F\downarrow a} \Psi_{F \uparrow a}.
\end{array}
\label{fesh20}
\end{equation}
Here, and below, we assume $u_0 < 0$, which is necessary for being near the Feshbach resonance.
Note that $\Psi_B$ couples to the fermions just like the boson field in the Bose-Fermi
model in (\ref{fesh7}), which is the reason for choosing this notation. If we perform the integral over
$\Psi_B$ in (\ref{fesh20}), we recover (\ref{fesh19}), as required.
For the large $N$ expansion, we have to integrate over $\Psi_{F \sigma a}$ first and obtain an effective
action for $\Psi_B$.
Because the action in (\ref{fesh20}) is Gaussian in the $\Psi_{F \sigma a}$, the integration 
over the fermion field involves evaluation of a functional determinant, and has the schematic form
\begin{equation}
\mathcal{Z}_{Fs} = \int {\cal D} \Psi_B (x, \tau)
\exp\left( -N \mathcal{S}_{\rm eff} \left[ \Psi_B (x, \tau) \right] \right),
\label{fesh21}
\end{equation}
where $\mathcal{S}_{\rm eff}$ is the logarithm of the fermion determinant of a single flavor. The key point
is that the only $N$ dependence is in the prefactor in (\ref{fesh21}), and so the theory of $\Psi_B$ can controlled
in powers of $1/N$. 

We can expand $\mathcal{S}_{\rm eff}$ in powers
of $\Psi_B$: the $p$'th term has a fermion loop with $p$ external $\Psi_B$ insertions. Details can be found in Refs.~\cite{predrag,veillette}. Here, we only note that the expansion to quadratic order at $\mu=\delta=T=0$, in which case
the co-efficient is precisely the inverse of the fermion $T$-matrix in (\ref{fesh2}):
\begin{equation}
\mathcal{S}_{\rm eff}  \left[ \Psi_B (x, \tau) \right] =  - \frac{1}{2} \int \frac{d \omega}{2 \pi} \frac{d^d k}{( 2\pi)^d} 
\frac{1}{T (k, i \omega) } |\Psi_B (k, \omega)|^2  + \ldots \label{fesh22}
\end{equation}
Given $S_{\rm eff}$, we then have to find its saddle point with respect to $\Psi_B$. At $T=0$, we will find the optimal
saddle point at a $\Psi_B \neq 0$ in the region of Fig.~\ref{fig:unitary} with a non-zero density: this means that the ground
state is always a superfluid of fermion pairs. The traditional expansion about this saddle point yields the $1/N$ expansion,
and many experimental observables have been computed in this manner \cite{predrag,veillette,veillette2}.

We conclude our discussion of the unitary Fermi gas by deriving an exact relationship between the total energy, $E$,
and the momentum distribution function, $n(k)$, of the fermions \cite{tan1,tan2}. 
We will do this using the structure of the large $N$ expansion.
However, we will drop the flavor index $a$ below, and quote results directly for the physical case of $N=2$.
As usual, we define the momentum distribution function by
\begin{equation}
n (k) = \langle \Psi_{F \sigma}^{\dagger} (k, t) \Psi_{F \sigma} (k, t) \rangle,
\label{fesh23}
\end{equation}
with no implied sum over the spin label $\sigma$. The Hamiltonian of the system in (\ref{fesh19}) is the
sum of kinetic and interaction energies: the kinetic energy is clearly an integral over $n(k)$ and so we can write
\begin{eqnarray}
E &=& 2 V \int \frac{d^d k}{( 2\pi)^d} \frac{k^2}{2m} n(k)  + u_0 V \langle \Psi_{F\uparrow}^\dagger \Psi_{F\downarrow}^\dagger
\Psi_{F\downarrow} \Psi_{F\uparrow} \rangle \nonumber \\
&=& 2 V \int \frac{d^d k}{( 2\pi)^d} \frac{k^2}{2m} n(k)  -  u_0 \frac{\partial \ln Z_{Fs}}{\partial u_0}. \label{fesh24}
\end{eqnarray}
where $V$ is the system volume, and all the $\Psi_F$ fields are at the same $x$ and $t$.  
Now let us evaluate the $u_0$ derivative using the expression for $Z_{Fs}$
in (\ref{fesh20}); this leads to 
\begin{equation}
\frac{E}{V} = 2 \int \frac{d^d k}{( 2\pi)^d} \frac{k^2}{2m} n(k) + \frac{1}{u_0} \left \langle \Psi^\ast_B (x, t) \Psi_B (x, t)  \right\rangle.
\label{fesh25}
\end{equation}
Now using the expression (\ref{fesh4}) relating $u_0$ to the scattering length $a$ in $d=3$, we can write this expression as
\begin{equation}
\frac{E}{V} = \frac{m}{4 \pi a} \left\langle \Psi_B^\ast \Psi_B \right \rangle + 2 \int \frac{d^3 k}{( 2\pi)^3} \frac{k^2}{2m} 
\left( n(k) - \frac{\left\langle \Psi_B^\ast \Psi_B \right \rangle m^2}{k^4} \right) \label{fesh26}
\end{equation}
This is the needed universal expression for the energy, expressed in terms of $n(k)$ and the scattering length,
and independent of the short distance structure of the interactions.

At this point, it is useful to introduce ``Tan's constant'' $C$, defined by \cite{tan1,tan2}
\begin{equation} 
C = \lim_{k \rightarrow \infty} k^4 n (k). \label{fesh27}
\end{equation}
The requirement that the momentum integral in (\ref{fesh26}) is convergent in the ultraviolet implies
that the limit in (\ref{fesh27}) exists, and further specifies its value
\begin{equation} 
C = m^2 \left\langle\Psi_B^\ast \Psi_B \right \rangle. \label{fesh28}
\end{equation}

We now note that the relationship $n (k) \rightarrow m^2  \left\langle\Psi_B^\ast \Psi_B \right \rangle/k^4$
at large $k$ is also as expected from a scaling perspective. We saw in Section~\ref{sec:fb} that the fermion
field $\Psi_F$ does not acquire any anomalous dimensions, and has scaling dimension $d/2$. Consequently
$n(k)$ has scaling dimension zero. Next, note that the operator $\Psi_B^\ast \Psi_B$ is conjugate to the
detuning from the Feshbach critical point; from (\ref{fesh18}) the detuning has scaling dimension $d-2$,
and so $\Psi_B^\ast \Psi_B$ has scaling dimension $d+z - (d-2) = 4$. Combining these scaling dimensions,
we explain the $k^{-4}$ dependence of $n(k)$.

\begin{figure}
\centerline{\includegraphics[width=2.5in]{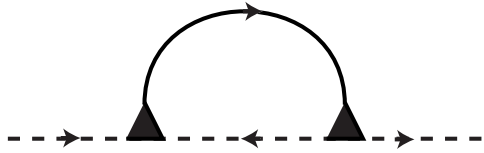}}
\caption{Order $1/N$ correction to the fermion Green's function.
Notation is as in Fig.~\ref{fig:fesh1}.} \label{fig:fesh3}
\end{figure}

It now remains to establish the claimed exact relationship in (\ref{fesh28}) as a general property
of a spinful Fermi gas near unitarity. As a start, we can examine the large $k$ limit of $n(k)$ in the BCS mean
field theory of the superfluid phase: the reader can easily verify that the text-book BCS expressions for
$n(k)$ do indeed satisfy (\ref{fesh28}). However, the claim of Refs.~\cite{braaten,sonunitary} is that (\ref{fesh28}) is exact 
beyond mean field theory, and also holds in the non-superfluid states at non-zero temperatures.
A general proof was given in Refs.~\cite{sonunitary}, and relied on the operator product expansion (OPE)
applied to the field theory (\ref{fesh20}). The OPE is a general method for describing the short
distance and time (or large momentum and frequency) behavior of field theories. Typically, in the Feynman
graph expansion of a correlator, the large momentum behavior is dominated by terms in which the external
momenta flow in only a few propagators, and the internal momentum integrals can be evaluated
after factoring out these favored propagators. For the present situation, let us consider the $1/N$ correction
to the fermion Green's function given by the diagram in Fig.~\ref{fig:fesh3}. 
Representing the bare fermion and boson
Green's functions by $G_F$ and $G_B$ respectively, Fig~\ref{fig:fesh3} evaluates to
\begin{equation}
G_F^2  (k, \omega)  \int \frac{d^d p}{(2 \pi)^d}  \frac{d \Omega}{2 \pi} G_B (p, \Omega) G_F (-k +p, - \omega + \Omega).
\label{fesh29}
\end{equation}
Here $G_B$ is the propagator of the boson action $\mathcal{S}_{\rm eff}$ specified by (\ref{fesh22}).
In the limit of large $k$  and $\omega$, the internal $p$ and $\Omega$ integrals are
dominated by $p$ and $\Omega$ much smaller than $k$ and $\omega$; so we can approximate (\ref{fesh29}) by
\begin{eqnarray}
&& G_F^2  (k, \omega) G_F (-k, - \omega) \int \frac{d^d p}{(2 \pi)^d}  \frac{d \Omega}{2 \pi} G_B (p, \Omega)  \nonumber \\
&&~~= G_F^2  (k, \omega) G_F (-k, - \omega) \left \langle \Psi_B^\ast \Psi_B \right \rangle.
\label{fesh30}
\end{eqnarray}
This analysis can now be extended to all orders in $1/N$. Among these higher order contributions 
are terms which contribute self energy
corrections to the boson propagator $G_B$ in (\ref{fesh30}): it is clear that these can be summed to replace
the bare $G_B$ in (\ref{fesh30}) by the exact $G_B$. Then the value of $\left \langle |\Psi_B |^2 \right \rangle$ in (\ref{fesh30})
also becomes the exact value. All remaining contributions can be shown \cite{sonunitary} to fall off faster at large
$k$ and $\omega$ than the terms in (\ref{fesh30}). So (\ref{fesh30}) is the exact leading contribution to the fermion Green's function
in the limit of large $k$ and $\omega$ after replacing $\left \langle |\Psi_B |^2 \right \rangle$ by its exact value. 
We can now integrate (\ref{fesh30}) over $\omega$ to obtain $n(k)$ at large
$k$. Actually the $\omega$ integral is precisely that in (\ref{fesh15}), which immediately yields the needed relation (\ref{fesh28}).

Similar analyses can be applied to determine the the spectral functions of other observables \cite{veillette2,punk,randeriarf,randeriarf2,combescot,braaten2,sonunitary}.

Determining of the specific value of Tan's constant requires numerical computations in the 
$1/N$ expansion of (\ref{fesh21}). From the scaling properties of the Feshbach resonant fixed point in $d=3$,
we can deduce the result obeys a scaling form similar to (\ref{xx10}):
\begin{equation}
C = (2 m T)^2 \Phi_C \left( \frac{\mu}{T} , \frac{ \nu}{\sqrt{2mT}} \right),
\label{fesh31}
\end{equation}
where $\Phi_C$ is a dimensionless universal function of its dimensionless arguments; note that the arguments represent
the axes of Fig.~\ref{fig:unitary}. The methods of Refs~\cite{predrag,veillette} can now
be applied to (\ref{fesh28}) to obtain numerical results for $\Phi_C$ in the $1/N$ expansion.
We illustrate this method here by determining $C$ to leading order in the $1/N$ expansion at $\mu=\nu =0$.
For this, we need to generalize the action (\ref{fesh22}) for $\Psi_B$ to $T>0$ and general $N$. Using (\ref{fesh2}) we can
modify (\ref{fesh22}) to
\begin{equation}
\mathcal{S}_{\rm eff} = N T \sum_{\omega_n} \int \frac{d^3 k}{8 \pi^3} \left[
D_0 (k, \omega_n) + D_1 (k, \omega_n) \right] |\Psi_B (k, \omega_n)|^2 ,
\label{fesh32}
\end{equation}
where $D_0$ is the $T=0$ contribution, and $D_1$ is the correction at $T>0$:
\begin{eqnarray}
&& D_0 (k, \omega_n) = \frac{m^{3/2}}{16 \pi} \sqrt{ - i \omega_n + \frac{k^2}{4 m}} \label{fesh33} \\
&& D_1(k, \omega_n) = \frac{1}{2} \int \frac{d^3 p}{8 \pi^3} \frac{1}{(e^{p^2/(2 mT)} + 1)} \frac{1}{\left( -i \omega + p^2/(2m)
+ (p+k)^2/(2m) \right)} . \nonumber
\end{eqnarray}
We now have to evaluate $\left\langle\Psi_B^\ast \Psi_B \right \rangle$ using the Gaussian action in (\ref{fesh32}).
It is useful to do this by separating the $D_0$ contribution, which allows us to properly deal with the large frequency
behavior. So we can write
\begin{equation}
\left\langle\Psi_B^\ast \Psi_B \right \rangle = \frac{1}{N} T \sum_{\omega_n} \int \frac{d^3 k}{8 \pi^3}
\left[ \frac{1}{D_0 (k, \omega_n) + D_1 (k, \omega_n)} - \frac{1}{D_0 (k, \omega_n)} \right]
+ D_{00}. \label{fesh34}
\end{equation}
In evaluating $D_{00}$ we have to use the usual time-splitting method to ensure that the bosons are normal-ordered,
and evaluate the frequency summation by analytically continuing to the real axis:
\begin{eqnarray}
D_{00} &=& \frac{1}{N}  \int \frac{d^3 k}{8 \pi^3} \lim_{\eta \rightarrow 0} 
T \sum_{\omega_n} \frac{e ^{i \omega_n \eta}}{D_0 (k, \omega_n)}  \nonumber \\
&=& \frac{16 \pi}{N m^{3/2}}  \int \frac{d^3 k}{8 \pi^3} \int_{\frac{k^2}{4m}}^{\infty} \frac{d \Omega}{\pi}
\frac{1}{(e^{\Omega/T} - 1)} \frac{1}{\sqrt{ \Omega - k^2 / (4 m)}}. \nonumber \\
&=& \frac{8.37758}{N}\, T^2  \label{fesh35}
\end{eqnarray}
The frequency summation in (\ref{fesh34}) can be evaluated directly on the imaginary frequency axis:
the series is convergent at large $\omega_n$, and is easily evaluated by a direct numerical summation.
Numerical evaulation of (\ref{fesh34})  now yields
\begin{equation}
C = (2mT)^2 \left( \frac{0.67987}{N} + \mathcal{O} (1/N^2) \right)
\end{equation}
at $\mu = \nu = 0$.

\end{document}